\documentclass[fleqn,usenatbib]{mnras}

\usepackage{newtxtext,newtxmath}

\usepackage{float}

\usepackage[T1]{fontenc}
\usepackage{ae,aecompl}

\usepackage{graphicx}	
\usepackage{amsmath}	
\usepackage{amssymb}	

\title[A hyperluminous obscured quasar at $z\approx4.3$]{A hyperluminous obscured quasar at a redshift of $z\approx4.3$}

\author[A.Efstathiou et al.]{\parbox{\linewidth}{Andreas Efstathiou$^{1}$\thanks{E-mail: a.efstathiou@euc.ac.cy}, Katarzyna Ma\l{}ek$^{2,3}$, Denis Burgarella$^3$, Peter Hurley$^4$, Seb Oliver$^4$, Veronique Buat$^3$, Raphael Shirley$^{4,5,6}$, Steven Duivenvoorden$^4$, Vicky\\ Papadopoulou Lesta$^1$, Duncan Farrah$^{7,8}$, Kenneth J. Duncan$^{9,10}$, María del Carmen Campos Varillas$^{4}$}
\\
\\
$^{1}$School of Sciences, European University Cyprus, Diogenes street, Engomi, 1516 Nicosia, Cyprus\\
$^{2}$National Centre for Nuclear Research, ul. Pasteura 7, 02-093 Warszawa, Poland\\
$^{3}$Aix Marseille Univ, CNRS, CNES, LAM Marseille, France\\
$^{4}$Astronomy Centre, Department of Physics and Astronomy, University of Sussex, Brighton, BN1 9QH, UK\\
$^{5}$Instituto de Astrof\'isica de Canarias, E-38205 La Laguna, Tenerife, Spain\\ $^{6}$Universidad de La Laguna, Dpto. Astrof\'isica, E-38206 La Laguna, Tenerife, Spain\\
$^{7}$Department of Physics and Astronomy, University of Hawaii, 2505 Correa Road, Honolulu, HI 96822, USA\\
$^{8}$Institute for Astronomy, 2680 Woodlawn Drive, University of Hawaii, Honolulu, HI 96822, USA\\
$^{9}$ Royal Observatory Edinburgh, University of Edinburgh, Blackford Hill, Edinburgh EH9 3HJ, UK\\
$^{10}$ Sterrewacht Leiden, Universiteit Leiden, Leiden, Netherlands
}

\date{Accepted 2020 December 21. Received 2020 December 21; in original form 2020 July 14}

\pubyear{2021}

\begin{document}
\label{firstpage}
\pagerange{\pageref{firstpage}--\pageref{lastpage}}
\maketitle

%
\begin{abstract}
In this work we report the discovery of the hyperluminous galaxy HELP\_J100156.75+022344.7 at the photometric redshift of $z \approx 4.3$. The galaxy was discovered in the Cosmological Evolution Survey (COSMOS) field, one of the fields studied by the Herschel Extragalactic Legacy Project (HELP). We present the spectral energy distribution (SED) of the galaxy and fit it with the CYprus models for Galaxies and their NUclear Spectra (CYGNUS) multi-component radiative transfer models. We find that its emission is dominated by an obscured quasar with a predicted total 1-1000$\mu m$ luminosity  of $3.91^{+1.69}_{-0.55} \times 10^{13} L_\odot$ and an active galactic nucleus (AGN) fraction of $\sim 89\%$. We also fit HELP\_J100156.75+022344.7 with the Code Investigating GALaxy Emission  (CIGALE) code and find a similar result.  This is only the second $z > 4$ hyperluminous obscured quasar discovered to date. The discovery of  HELP\_J100156.75+022344.7 in the $\sim 2$deg$^2$ COSMOS field implies that a large number of obscured hyperluminous quasars may lie in the HELP fields which cover $\sim 1300$deg$^2$. If this is confirmed, tension between supermassive black hole evolution models and observations will be alleviated. We estimate the space density of objects like HELP\_J100156.75+022344.7 at $z \approx 4.5$ to be $\sim 1.8 \times 10^{-8}$Mpc$^{-3}$. This is slightly higher than the space density of coeval hyperluminous optically selected quasars suggesting that the obscuring torus in $z > 4$ quasars may have a covering factor $\gtrsim 50\%$. 

\end{abstract}




\vspace{-60pt}
\section{Introduction}

Ever since the discovery of quasars in the 1960s, numerous surveys searching for quasars of ever increasing redshift have been carried out.  The main motivation of the surveys was to take advantage of the extreme luminosity of quasars to search for distant objects that probe the conditions in the early history of the Universe. Luminous quasars selected at optical and near-infrared wavelengths have now been observed out to $z > 6$ \citep[e.g.][]{ wu15} corresponding to an epoch less than a billion years after the big bang. Currently, just a handful of quasars have also been discovered at $z > 7$ \citep*[e.g.][]{ mortlock11, banados18, wang18, yang20}.

The observed space density of quasars rises steeply from $z \sim 6-7$ reaching a peak at $z \sim 2$ \citep[e.g.][]{ kelly10} and then declines again towards $z \sim 0$. In the redshift range $0-3$ the black hole accretion rate density (BHAD) and the star formation rate density (SFRD) correlate \citep[e.g.][]{ delvecchio14} and this is one of
 the reasons why accretion of matter onto supermassive black holes (SMBH) in quasars, or more generally active galactic nuclei (AGN), and star formation are believed to be inter-related phenomena \citep*[e.g.][]{ fabian12, hickox18}. At $z > 3$, however, a large deviation between SFRD and BHAD is observed with BHAD being much lower than predicted by SMBH evolution models \citep{vito18}.

The deviation between BHAD and SFRD may be due to the fact that there are large numbers of obscured quasars at $z > 3$ which have not yet been discovered. This is expected from the unified model for AGN \citep{antonucci93} if it holds at high-redshift.  The model postulates the existence of a geometrically and optically thick dusty torus, an idea which is well tested in the local universe, which obscures the central engine and reprocesses its optical and ultraviolet radiation to predominantly 3-30$\mu m$ radiation. However, the mean covering factor of the torus which determines the proportion of quasars which are obscured may depend on redshift as well as luminosity \citep[e.g.][]{ hickox18}. Detecting obscured AGN at $z > 3$ is made difficult by the fact that the peak of the emission from the torus shifts to the far-infrared. X-ray telescopes also currently lack the sensitivity to find obscured AGN at $z > 3$ unless they are extremely luminous.

The Wide-field Infrared Survey Explorer (WISE) carried out a survey of the whole sky at mid-infrared wavelengths and detected a large number of objects which are especially luminous at those wavelengths \citep*{eisenhardt12, bridge13}. These objects have been named Hot Dust Obscured Galaxies (Hot DOGs) and are believed to be powered by emission by an obscured AGN \citep*[e.g.][]{ jones15, farrah17}.  However, to date only one $z > 4$ obscured AGN has been found by WISE, the Extremely Luminous Infrared Galaxy W2246-0526 at $z=4.6$ with a luminosity of $\sim 3 \times 10^{14} L_\odot$ \citep*{tsai15, tsai18} making it the most luminous galaxy observed to date.  \citet{diaz-santos18} recently imaged W2246-0526 with ALMA and found it to be a multiple merger.

The Herschel Extragalactic Legacy Project \citep[HELP;][]{ shirley19} which assembled ultraviolet to submillimetre spectral energy distributions for over 170 million galaxies in about 1300 deg$^2$ of the sky is much deeper than WISE. Early results from HELP have been presented in \citet{malek18}. The HELP database, which is already  public, is therefore much more suitable for determining the space density of $z > 4$ obscured quasars and exploring the physics of AGN and their role in galaxy evolution at those redshifts.  

In this Letter we report the discovery of an obscured AGN in the COSMOS field \citep{scoville07} with an estimated photometric redshift of 4.48 \citep{laigle16} which was discovered as part of the HELP project. The catalogue name of the source is HELP\_J100156.758+022344.739 but in the rest of the letter we will refer to it as HELP\_J100156.75+022344.7.  We use the  CYprus models for Galaxies and their NUclear Spectra (CYGNUS)\footnote{available at http://ahpc.euc.ac.cy/} multi-component radiative transfer models for AGN tori, starbursts and spheroidal hosts to fit the spectral energy distribution (SED) of HELP\_J100156.75+022344.7 with the MCMC code SATMC \citep{johnson13}. Using SATMC we also simultaneously obtain a photometric redshift using all of the multi-wavelength data used in the SED fit. For comparison we also fit the SED of HELP\_J100156.75+022344.7 with the CIGALE code \citep*{noll09, boquien19} and find similar results.

In section 2 of this Letter we discuss the method of assembling the multi-wavelength observations, in section 3 we discuss the radiative transfer models we used and the method of fitting the SED of HELP\_J100156.75+022344.7 and in section 4 we discuss our results. We assume a standard cosmology with $H_0 = 70$ km s$^{-1}$ Mpc$^{-1}$, $\Omega =1$ and $\Lambda = 0.7$.

\vspace{-15pt}
\section{Observations}

\begin{table}
	\centering
\caption{Photometry for HELP\_J100156.75+022344.7 from the optical to the submillimetre. The errors in the Herschel fluxes include the errors associated with the deblending.}
	\label{tab:example_table}
	\begin{tabular}{cccc} 
		\hline
		Wavelength &  Flux Density  &  Filter \\
		  $\lambda$ ($\mu m$)  &     $S_\nu$ (Jy)  & \\
		\hline

0.446  &    0.00 $\pm$ 1.10 $\times~ 10^{-8}$  &    SUBARU B$^\dagger$ \\  
0.548  &    0.00 $\pm$ 2.10 $\times~ 10^{-8}$   &   SUBARU V$^\dagger$ \\   
0.628  &    4.71 $\pm$ 0.95 $\times~ 10^{-8}$   &   MEGACAM R \\    
0.650  &    1.25 $\pm$ 0.17 $\times~ 10^{-7}$    &   SUPRIME RC \\    
0.761  &    1.21 $\pm$ 0.13 $\times~ 10^{-7}$   &   MEGACAM I  \\  
0.767  &    1.57 $\pm$ 0.24 $\times~ 10^{-7}$   &   SUPRIME IP \\  
0.873  &    1.65 $\pm$ 0.30 $\times~ 10^{-7}$   &   SUPRIME Z  \\    
0.921  &    5.31 $\pm$ 0.65 $\times~ 10^{-7}$   &   SUPRIME N921 \\    
1.645  &    3.63 $\pm$ 0.57 $\times~ 10^{-7}$   &   VISTA H  \\       
2.147  &    9.47 $\pm$ 0.55 $\times~ 10^{-7}$   &   VISTA Ks \\ 
3.54   &    2.96 $\pm$ 0.06 $\times~ 10^{-6}$   &   IRAC 1  \\       
4.48   &    2.39 $\pm$ 0.05 $\times~ 10^{-6}$   &   IRAC 2  \\        
5.7    &    2.13 $\pm$ 1.50 $\times~ 10^{-6}$   &   IRAC 3$^\dagger$  \\      
7.83   &    9.63 $\pm$ 2.27 $\times~ 10^{-6}$   &   IRAC 4 \\  
24     &    6.50 $\pm$ 0.05 $\times~ 10^{-4}$   &   MIPS24 \\  
70     &    0.00 $\pm$ 3.00 $\times~ 10^{-3}$   &  MIPS70$^\dagger$ \\  
100   &     5.15 $\pm$ 1.12 $\times~ 10^{-3}$   &    PACS \\        
160   &     1.01 $\pm$ 1.00 $\times~ 10^{-2}$   &   PACS$^\dagger$ \\    
250   &     9.11 $\pm$ 0.93 $\times~ 10^{-3}$   &   SPIRE \\                 
350   &     1.03 $\pm$ 0.12 $\times~ 10^{-2}$   &   SPIRE \\  
500   &     9.03 $\pm$ 1.31 $\times~ 10^{-3}$   &   SPIRE \\

		\hline
	\end{tabular}
	
	  $^\dagger$ Upper limits

\end{table}

 HELP uses prior information from optical, near-infrared and mid-infrared surveys to deal with blending in the confusion limited maps from the Herschel Space Observatory \citep{pilbratt10}. \citet{hurley17} developed the XID+ method of deblending confused Herschel sources. One of the first fields analyzed with XID+ was the well studied COSMOS field \citep{scoville07} in which we identified HELP\_J100156.75+022344.7. \citet{shirley19} describe in detail the process of constructing the HELP catalogue from pristine catalogues produced by other independent teams. The HELP photometry for HELP\_J100156.75+022344.7 we use in the rest of the Letter, and which is the result of the standard default processing by XID+, is listed in Table 1. The source is very faint in the optical and this partly explains why a spectrum is not available. For example the AB magnitude in the SUPRIME RC filter is 26.2. The source is not detected in the SUBARU B and V filters \citep{laigle16}. We use these data as upper limits in our analysis. The photometric uncertainties of the SPIRE data-points are relatively small as HELP\_J100156.75+022344.7 is fairly isolated in the SPIRE images. This is shown in Figure 1 where all the detections are plotted and in Figure S4 where the p-value maps derived by XID+ are plotted.
  We have checked if HELP\_J100156.75+022344.7 is included in the 70$\mu m$ catalogue of \citet{frayer09}. The nearest 70$\mu m$ source is 1.27’ away from HELP\_J100156.75+022344.7. We have used the noise of the Frayer et al map at the approximate position of HELP\_J100156.75+022344.7 to set an upper limit at 70$\mu m$ which we use in the SED fit. HELP\_J100156.75+022344.7 is detected in the VLA-COSMOS 3 GHz catalogue of \citet{smolcic17} who quote a flux of $27 \pm 2.6\mu$Jy and classify the source as a moderate-to-high radiative luminosity AGN. There are no detections in the Chandra catalogue of \citet{marchesi16} and the XMM-COSMOS catalogue of \citet{cappelluti09}. The source is also very faint in the HST F814W image. In Figure 1 we show postage stamps of the source in eighteen different filters. 
 
\begin{figure}
\centering
 \includegraphics[width=0.45\textwidth]{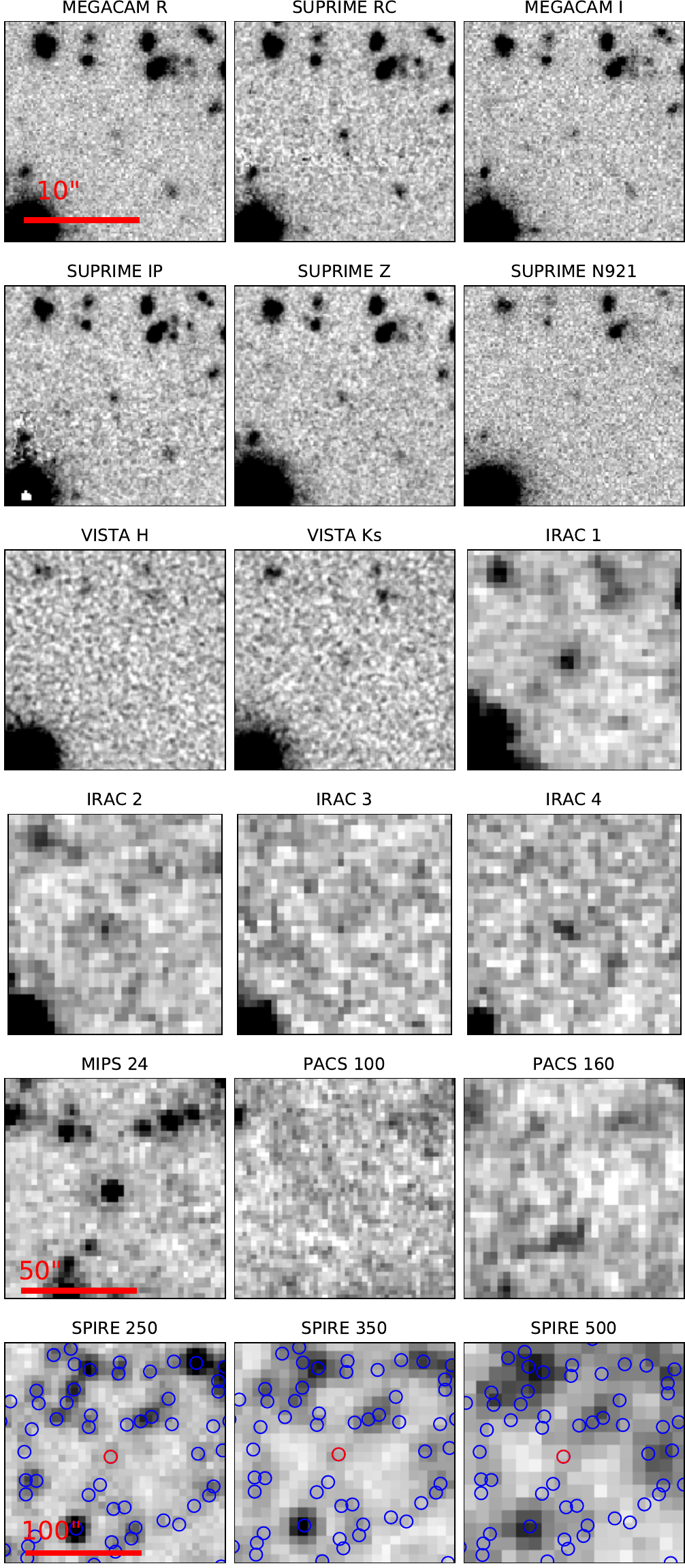}
 \vspace{-5pt}
 \caption{Postage stamps of HELP\_J100156.75+022344.7 in all the filters listed in Table 1 and used in the SED fit except MIPS70 and SUBARU B and V which are not included in HELP.  In the last two rows we have a zoom out by factors of 5 and 10. The blue circles denote all the detections by XID+ and the red circles the source.}
    \label{fig:example_figure}
     \vspace{-15pt}
\end{figure}

\begin{table}
	\centering
\caption{Derived physical quantities and their 1$\sigma $ uncertainties for HELP\_J100156.75+022344.7 by the CYGNUS models and CIGALE. All listed luminosities are 1-1000$\mu m$ luminosities. The starburst SFR is averaged over 10Myr in both CYGNUS and CIGALE. The minimum reduced $\chi^2$ of the best fits with CYGNUS and CIGALE are 4.3 and 4.8 respectively.}
	\label{tab:example_table}
	\begin{tabular}{llll} 
		\hline
		Quantity &  CYGNUS   &  CIGALE  &  Unit \\
		\hline
     
  AGN Luminosity   & ${   3.47^{+1.79}_{-0.58}}$    & $1.27^{+0.07}_{-0.07}$  & $10^{13}$  L$_\odot$   \\  
  Starburst Luminosity & ${   4.34^{+1.12}_{-0.53}}$    & -  & $10^{12}$ L$_\odot$ \\   
  Spheroidal Luminosity          & ${   9.54^{+3.27}_{-1.15}}$    & -  & $10^{10}$ L$_\odot$  \\ 
Total Luminosity     & ${   3.91^{+1.69}_{-0.55}}$     & $1.72^{+0.06}_{-0.06}$ &  $10^{13}$ L$_\odot$  \\ 
Starburst SFR      & ${   1040^{+110}_{-269}}$   & -  &  M$_\odot$ yr$^{-1}$\\  
Spheroidal SFR      & ${   12^{+4}_{-2}}$    & - & M$_\odot$ yr$^{-1}$\\  
Total SFR      & ${   1051^{+109}_{-268}}$    & $991^{+223}_{-223}$   & M$_\odot$ yr$^{-1}$ \\ 
Starburst Stellar Mass      & ${   1.00^{+0.11}_{-0.26}}$     & - &  $10^{10}$ M$_\odot$\\  
Spheroidal Stellar Mass      & ${   4.03^{+1.04}_{-0.49}}$     & - &  $10^{10}$ M$_\odot$ \\  
Total Stellar Mass     & ${   5.03^{+0.96}_{-0.45}}$     &  $7.55^{+1.15}_{-1.15}$   & $10^{10}$ M$_\odot$ \\  
Core-collapse SN rate   & ${   2.63^{+0.29}_{-0.68}}$            & -  & SN yr$^{-1}$  \\ 
    AGN fraction        & ${   0.89^{+0.05}_{-0.03}}$  & $0.74^{+0.04}_{-0.04}$ \\ 

		\hline
	\end{tabular}
\end{table}

\vspace{-15pt}
\section{SED fitting with the CYGNUS radiative transfer models}
\label{sec:RT_models}

In this section we describe the libraries of spheroidal galaxy, AGN torus and starburst models used for the SED fitting of HELP\_J100156.75+022344.7 with SATMC \citep{johnson13}. The models form part of the CYGNUS\footnote{available at http://ahpc.euc.ac.cy/}  collection of radiative transfer models. Various combinations of these models have been used extensively to interpret the observed SEDs of a broad range of galaxies in \citet*{alexander99, ruiz01, farrah02, verma02, farrah03, farrah12, mattila12, efstathiou13, lonsdale15, harris16, farrah17, herrero-illana17, mattila18, pitchford19}.

\vspace{-15pt}
\subsection{Spheroidal models} 
\label{sec:RT_models_sph}

The method employed for computing the libraries of spheroidal models used in this paper is an evolution of the cirrus model of \citet{efstathiou03}.  As in \citet{efstathiou03}, the models of \citet*{bruzual93, bruzual03} are used in combination with an assumed star formation history (SFH) to compute the spectrum of starlight which is illuminating the dust throughout the model galaxy.  \citet{efstathiou03} assumed an exponentially decaying star formation rate $\dot{M}_{\ast}$ whereas here we assume a more general delayed exponential ($\dot{M}_{\ast} \propto t~e^{-{{t}\over {\tau^s}}}$), where $t$ is the time since the big bang and $\tau^s$ is the e-folding time of the exponential.

An important new feature of the spheroidal model is that the stars and dust are assumed to be mixed in a distribution that assumes a S\'{e}rsic profile with $n = 4$. A spherical geometry was chosen for the distribution of stars and dust hence the name `spheroidal'. The dust model used here is the same as that used in \citet{efstathiou09} and the models are computed with an adapted version of the spherically symmetric radiative transfer code used in the same paper. 

The spheroidal model assumes {three} parameters: the e-folding time of the delayed exponential $\tau^s$, the optical depth of the spherical cloud from its centre to its surface $\tau_v^s$ and the ratio of the central stellar emission to that of the intensity of starlight in the solar neighborhood $\psi^s$.   The library used in this paper was computed at a redshift of $z=4.4$. We assumed that all the stars in the galaxy formed with a Salpeter IMF out of gas with a metallicity of 5\% of solar, which appears to be typical of gas at such redshifts \citep[e.g.][]{ rafelski12}.

\vspace{-15pt}
\subsection{AGN torus and starburst models}

We use the library of AGN torus models computed with the method of \citet{efstathiou95} and described in more detail in \citet{efstathiou13}. The method models the torus predicted by the AGN unified scheme with a smooth tapered disk distribution. The tapered disk consists of multiple species of dust of various sizes and compositions and its density declines as $r^{-1}$ where $r$ is the distance from the central SMBH. 

The AGN model parameters and their assumed range are the half-opening angle of the torus ($\Theta_0 = 30\degr - 75\degr$), the inclination of the torus ($i = 0\degr - 90\degr$), the ratio of inner to outer disc radius ($r_1/r_2 = 0.01 - 0.05$)  and the equatorial optical depth at 1000\AA ~~($\tau_{uv} = 250 - 1250$; in the dust model used in this paper this translates to an $A_V\approx \tau_{uv}/5$ and $\tau_{9.7\mu m} \approx \tau_{uv}/61$).


We use the starburst model originally developed by \citet{efstathiou00} and updated by \citet{efstathiou09}. The model incorporates the stellar population synthesis model of \citep*{bruzual93, bruzual03}. The starburst model parameters and their assumed range are: the age of the starburst $t_* = 0-30Myr$, the giant molecular clouds (GMC's) initial optical depth $\tau_V=50-250$,  the e-folding time of the exponentially decaying star formation rate (SFR) $\tau=10-30Myr$. 

\vspace{-15pt}
\subsection{SED fitting and photo-z estimating with SATMC}

We fitted the SED of HELP\_J100156.75+022344.7 with the MCMC code SATMC \citep{johnson13}. SATMC can fit an SED given libraries of radiative transfer models and provides the option to simultaneously determine a photometric redshift which takes into account all the multi-wavelength data used in the fit. 

SATMC predicts a photo-z of $4.71^{+0.76}_{-0.56}$ for HELP\_J100156.75+022344.7 which agrees well with the photometric redshift  estimates of 4.48 and 4.08 determined by \citet{laigle16} and \citet{chang17} respectively. We note that in the SUPRIME N921 filter we see an excess in the photometry which deviates significantly from the fit. If this is due to an emission line it would need to be centered at around 1727\AA~~ assuming a redshift of 4.33, the value given in the HELP catalogue. The nearest strong quasar emission line is CIII] with a rest wavelength of 1909\AA. If the excess is due to the CIII] line this would imply the actual redshift of HELP\_J100156.75+022344.7 is close to 3.82. Alternatively there could be a problem with the  SUPRIME N921 photometry.

In Figure 2 we show the best fit of the SED obtained by fixing the redshift at 4.33. As the far-infrared and submillimetre photometry is very limited we also fix the starburst age $t_*$ at 10Myr, the e-folding time of the starburst $\tau$ at 20Myr and $\psi^s$ at 5. We have a total of 10 free parameters in the fit:  $\tau^s$, $\tau_v^s$, $\Theta_0$, $i$, $r_1/r_2$, $\tau_{uv}$, $\tau_V$, $f_{AGN}$, $f_s$, $f_{SB}$. The last three are scaling factors that determine the luminosities of the AGN, spheroidal and starburst components respectively. The fit with SATMC is generally very good apart from the discrepancies in the SUPRIME N921 filter and IRAC4. The discrepancy in IRAC4 may suggest the presence of polar dust \citep*{efstathiou06, efstathiou13}.

SATMC determines and stores the best-fit parameters of the models and their likelihoods. Each combination of model parameters stored by SATMC has been post-processed using our own routines to extract the physical quantities of HELP\_J100156.75+022344.7 such as stellar mass, star formation rate, AGN fraction and their uncertainties and these are given in Table 2. We find in particular that HELP\_J100156.75+022344.7 is an AGN-dominated object with 89\% of its 1-1000$\mu m$ luminosity of $3.91 \times 10^{13} L_\odot$ provided by an obscured AGN. The co-variances between the fitted parameters of both the photo-z fit and the fit with fixed $z$ are plotted in Figures S2 and S3 respectively.

\begin{figure}
	\centering
	\vspace{-35pt}
	\includegraphics[width=0.75\textwidth, angle=180]{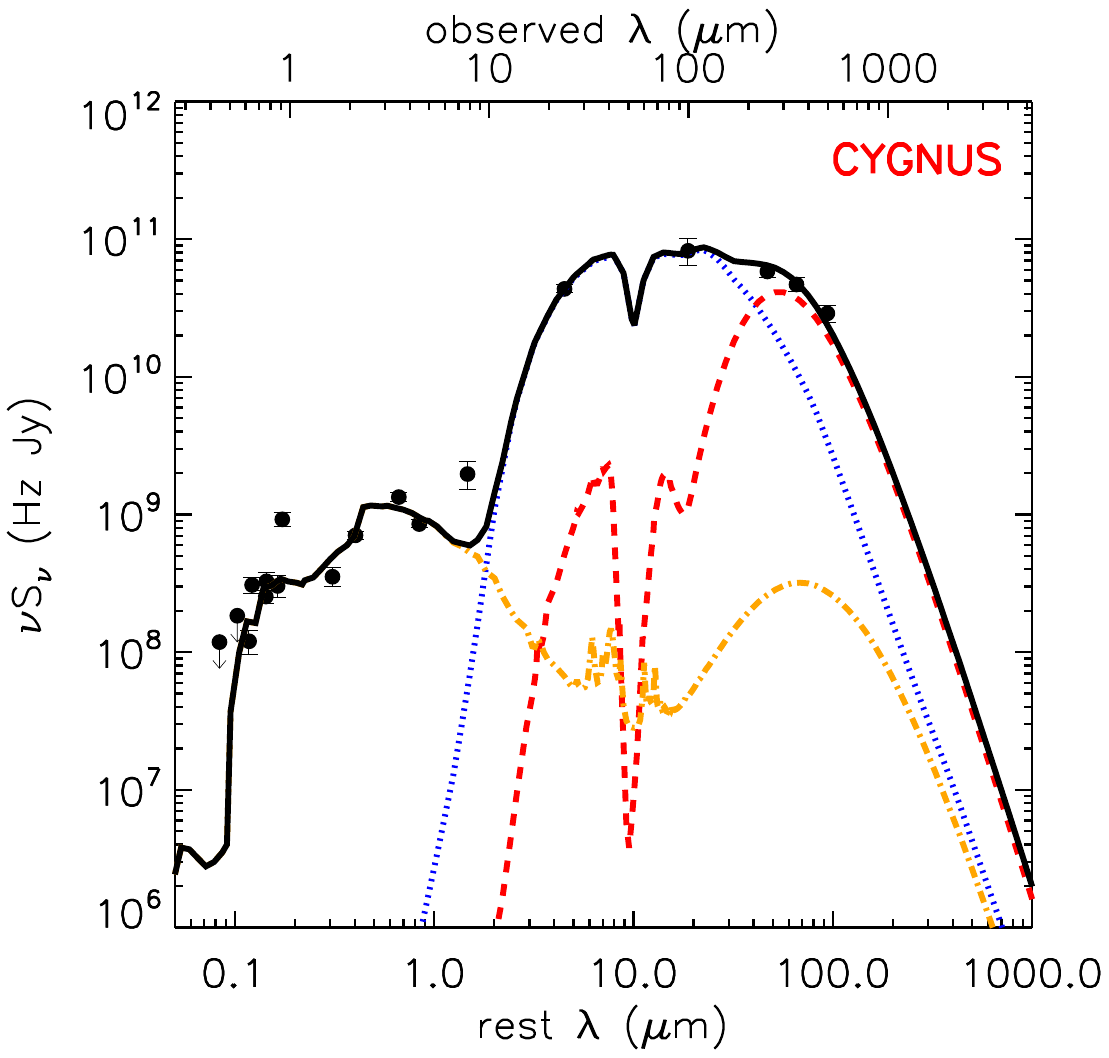}
	\vspace{-35pt}
    \caption{SED fit of HELP\_J100156.75+022344.7 with CYGNUS models using libraries of starburst (red dashed), AGN torus (blue dotted) and spheroidal models (orange dot-dashed). The total emission is indicated by the solid black line.}
    \label{fig:example_figure}
    \vspace{-15pt}
\end{figure}

\vspace{-15pt}
\section{Discussion}

In order to get an independent estimate of the main physical quantities we determined for  HELP\_J100156.75+022344.7 with CYGNUS, we also fitted the SED with the CIGALE code \citep*{noll09, boquien19}. 
The main physical quantities extracted with CIGALE are tabulated in Table 2.
This object was included in one of the HELP fields and was automatically fitted with the other galaxies in the COSMOS field. The standard HELP fitting procedure, described in detail in \citet{malek18},  estimated an AGN fraction of $80 \pm 4\%$. We performed dedicated SED fitting for this object using the CIGALE tool. We used the same modules as described in \citet{malek18}, but for the purposes of the comparison in this letter, we used a Salpeter IMF, and a metallicity of 0.004 (2\% of solar).  For the HELP project in general, we used templates from the smooth models of \citet{fritz06} to model the AGN torus emission. Here we use the same module but with a denser grid of possible parameters of AGN fraction and angle between the equatorial axis and line of sight to constrain these parameters more accurately. The dedicated CIGALE fit shown in Figure 4 predicts an AGN fraction of $74 \pm 4\%$.
The difference with the estimate from CYGNUS is most probably due to the $\sim {128}\%$ anisotropy correction applied to the luminosity of the AGN torus of \citet{efstathiou95}, see also \citet{efstathiou06}. This correction is not applied in the AGN module in CIGALE. In any case, it should be noted that the agreement between the two results is remarkable given the limited photometry and the significant differences in the methods used.

In Figure 3 we compare the SED of HELP\_J100156.75+022344.7 with that of the local deeply obscured ultraluminous infrared galaxy IRAS 08572+3915 (\citet{efstathiou14}, $L \approx 10^{13} L_\odot$), the $z=2.74$ Hot DOG WISE233759.50+792654.6 (\citet{farrah17}, $L \approx 10^{14} L_\odot$) and the $z=4.05$ hyperluminous galaxy GN20 (\citet{daddi09}, $L \approx 3 \times 10^{13} L_\odot$). According to the model of \citet{efstathiou14}, IRAS 08572+3915 is a local AGN-dominated galaxy which is estimated to have an AGN fraction of about 90\%. IRAS 08572+3915 is a much more obscured object compared to HELP\_J100156.75+022344.7 with a very deep silicate absorption feature as its torus is more optically thick. For comparison, our CYGNUS fit predicts $\tau_{uv} =  {261^{+66}_{-11}}$ and $i = {88^{+0.9}_{-0.7}}\degr$ for HELP\_J100156.75+022344.7 whereas the corresponding values for IRAS 08572+3915 from the \citet{efstathiou14} fit were $\tau_{uv} = 500$ and $i = 88\degr$.  If there are objects similar to IRAS 08572+3915 at $z > 4$ they would generally be missed by current surveys such as WISE and HELP because they are much fainter in the rest-frame mid-infrared. WISE233759.50+792654.6 with a predicted AGN fraction of about 90\% \citep{farrah17} has a very similar SED to HELP\_J100156.75+022344.7. 

The fact that the ultraviolet to far-infrared SED of HELP\_J100156.75+022344.7 can be fitted very well with two independent methods and that the SED is similar to known lower redshift Hot DOGs gives us confidence that there is no contribution by a foreground galaxy in the spectrum. We therefore consider gravitational amplification unlikely.

As expected for an object which is dominated by star formation \citep{riechers14}, GN20 is fainter in the rest-frame mid-infrared compared to HELP\_J100156.75+022344.7 by about an order of magnitude. \citet{spinoglio17} discussed the potential of a deep survey with SPICA at 34$\mu m$ to detect infrared galaxies up to redshift 6 assuming that such objects will have an SED similar to GN20. Figure 3 suggests that objects like HELP\_J100156.75+022344.7, IRAS 08572+3915 and WISE233759.50+792654.6 will be much easier to detect and study with the SPICA and JWST spectrometers compared to objects like GN20 if they exist at $z=4-6$.  

 \begin{figure}
 \centering
 	\vspace{-35pt}
	\includegraphics[width=0.75\textwidth, angle=180]{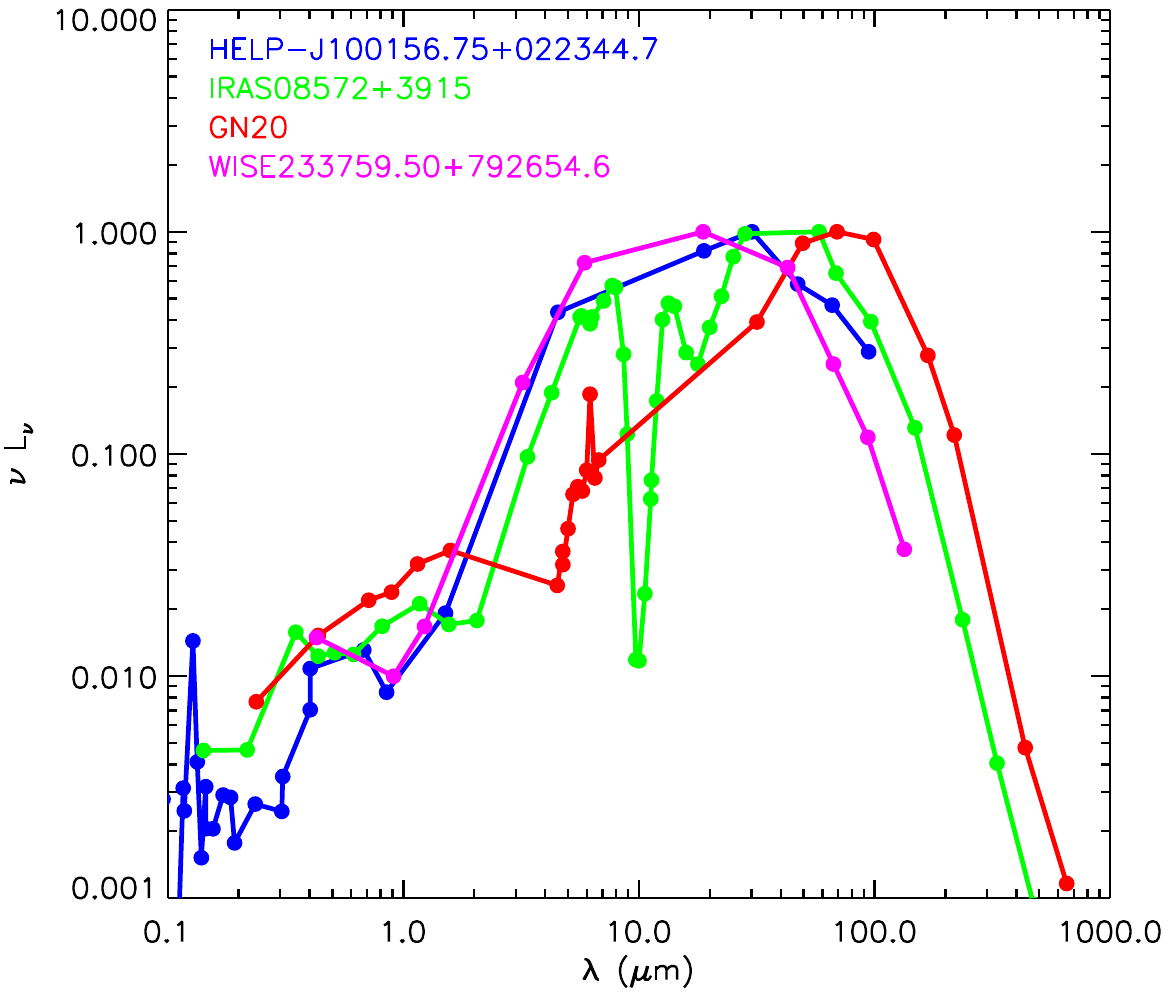}
    \vspace{-30pt}
    \caption{Comparison of the rest frame SEDs of GN20 (red), IRAS08572+3915 (green), WISE233759.50+792654.6 (magenta) and HELP\_J100156.75+022344.7 (blue). The SEDs have been normalized to their maximum $\nu L_{\nu}$.}
    \label{fig:example_figure}
     \vspace{-15pt}
\end{figure}

The space density of optically selected quasars at $z=4-5$ with M$_{1450, AB} = -27$, which corresponds to a bolometric luminosity of a  few times $10^{13} L_\odot$, is about $10^{-8} $Mpc$^{-3}$ \citep{akiyama18}. We estimate the space density of objects like the one we have discovered as follows: The comoving volume between $z=3-6$ for COSMOS, which we assume it has an area of $1.7 $deg$^2$,  is $5.53 \times 10^7$ Mpc$^3$. This gives a space density of  $1.8 \times 10^{-8} $Mpc$^{-3}$ which is slightly higher than the space density of optically selected quasars. This in turn implies that the covering factor of the obscuring torus in these objects is $\gtrsim 50\%$. This agrees well with the covering factor estimated from the Hot DOGs discovered by WISE \citep{assef15} and the dust reddened quasars in FIRST and UKIDDS \citep{glikman13}.

It is also interesting to compare the space density of obscured quasars at $z=4-6$ with the observed space density of red sources in the redshift interval $4-6$ which \citet{duivenvoorden18} estimate to be $1.1 \times 10^{-8}$ Mpc$^{-3}$. Extreme starbursts and extreme AGN are probably short-lived phenomena with similar duration so estimates of their space density, once refined by analyzing the complete HELP database, will give strong constraints about the co-evolution of extreme AGN and starbursts at those redshifts.

	\vspace{-15pt}
	\section*{Acknowledgements}

 We would like to thank an anonymous referee for useful comments and suggestions. HELP and the work leading to this paper has received funding from the European Union Seventh Framework Programme FP7/2007-2013/ under grant agreement no.607254.  AE, DF and VPL acknowledge support from the project EXCELLENCE/1216/0207/ GRATOS funded by the Cyprus RIF. KM has been supported by the National Science Centre (grant UMO-2018/30/E/ST9/00082).

\vspace{-15pt}
\section*{Data Availability Statement}

The data underlying this article are available in the article.





\vspace{-15pt}


\begin{thebibliography}{}


\bibitem[Akiyama et al. (2018)]{akiyama18}Akiyama, M., et al. 2018, PASJ, 70S, 44A

\bibitem[Assef et al. (2015)]{assef15}Assef, R.J., et al. 2015, ApJ, 804, 17 

\bibitem[Alexander et al.(1999)]{alexander99} 
Alexander, D.~M., et al.\ 1999, \mnras, 310, 78 

\bibitem[Antonucci (1993)]{antonucci93}Antonucci, R., 1993, ARAA, 31, 473  

\bibitem[Banados et al. (2018)]{banados18}Banados, E., et al. 2018, Nature, 553, 473 

\bibitem[Boquien et al.(2019)]{boquien19} 
Boquien, M., Burgarella, D., Roehlly, Y., et al.\ 2019, \aap, 622, A103 

\bibitem[Bridge et al.(2013)]{bridge13} 
Bridge, C.~R., Blain, A., Borys, C.~J.~K., et al.\ 2013, \apj, 769, 91 

\bibitem[Bruzual \& Charlot (1993)]{bruzual93} 
Bruzual,  G. \& Charlot, S.\ 1993, \apj, 405, 538 

\bibitem[Bruzual \& Charlot (2003)]{bruzual03} 
Bruzual, G. \& Charlot, S.\ 2003, \mnras, 344, 1000

\bibitem[Cappelluti et al. (2009)]{cappelluti09}Cappelluti, N., et al. 2009, AA, 497, 635 

\bibitem[Chang et al. (2017)]{chang17}Chang, Y. et al. 2017, ApJS, 33, 19

\bibitem[Daddi et al. (2009)]{daddi09}Daddi, E., et al. 2009, ApJ, 694, 1517  

\bibitem[Delvecchio et al. (2014)]{delvecchio14}Delvecchio, I., et al. 2014, MNRAS, 439, 2736  

\bibitem[Diaz-Santos et al. (2018)]{diaz-santos18}Diaz-Santos, T., et al. 2018, Science, 362, 1034   

\bibitem[Duivenvoorden et al. (2018)]{duivenvoorden18}Duivenvoorden, S., et al. 2018, MNRAS, 477, 1099  

\bibitem[Efstathiou (2006)]{efstathiou06} 
Efstathiou A.\ 2006, \mnras, 371, L70

\bibitem[Efstathiou \& Rowan-Robinson (1995)]{efstathiou95} 
Efstathiou A., Rowan-Robinson M.\ 1995, \mnras, 273, 649

\bibitem[Efstathiou, Rowan-Robinson \& Siebenmorgen (2000)]{efstathiou00} 
Efstathiou A., Rowan-Robinson M., Siebenmorgen R.\ 2000, \mnras, 313, 734

\bibitem[Efstathiou \& Rowan-Robinson (2003)]{efstathiou03} 
Efstathiou, A. \& Rowan-Robinson, M.\ 2003, \mnras, 343, 322 

\bibitem[Efstathiou \& Siebenmorgen (2009)]{efstathiou09} 
Efstathiou A., Siebenmorgen R.\ 2009, \aap, 502, 541

\bibitem[Efstathiou et al.(2013)]{efstathiou13} 
Efstathiou A., Christopher N., Verma A., Siebenmorgen R.\ 2013, \mnras, 436, 1873

\bibitem[Efstathiou et al.(2014)]{efstathiou14} 
Efstathiou, A., Pearson, C., Farrah, D., et al.\ 2014, \mnras, 437, L16 

\bibitem[Eisenhardt et al.(2012)]{eisenhardt12} 
Eisenhardt, P.~R.~M., Wu, J., Tsai, C.-W., et al.\ 2012, \apj, 755, 173 

\bibitem[Fabian (2012)]{fabian12}Fabian, A. C., 2012, ARAA, 50, 455 

\bibitem[Farrah et al.(2002)]{farrah02} 
Farrah, D., Serjeant, S., Efstathiou, A., et al.\ 2002, \mnras, 335, 1163

\bibitem[Farrah et al.(2003)]{farrah03} 
Farrah, D., Afonso, J., Efstathiou, A., et al.\ 2003, \mnras, 343, 585

\bibitem[Farrah et al.(2012)]{farrah12}
Farrah D. et al.\ 2012, \apj, 745, 178

\bibitem[Farrah et al.(2017)]{farrah17} 
Farrah, D., Petty, S., Connolly, B., et al.\ 2017, \apj, 844, 106

\bibitem[Frayer et al. (2009)]{frayer09}Frayer, D.T., et al. 2009, AJ, 138, 1261 

\bibitem[Fritz et al.(2006)]{fritz06} 
Fritz, J., Franceschini, A., \& Hatziminaoglou, E.\ 2006, \mnras, 366, 767

\bibitem[Glikman et al. (2013)]{glikman13}Glikman, E., et al. 2013, ApJ, 778, 117 

\bibitem[Harris et al.(2016)]{harris16} 
Harris, K., Farrah, D., Schulz, B., et al.\ 2016, \mnras, 457, 4179 

\bibitem[Herrero-Illana et al.(2017)]{herrero-illana17} 
Herrero-Illana, R., P{\'e}rez-Torres, M. {\'A}., Randriamanakoto, Z., et al.\ 2017, \mnras, 471, 1634 

\bibitem[Hickox \& Alexander (2018)]{hickox18}Hickox, R.C. \& Alexander, D.M, 2018, ARAA, 52, 625

\bibitem[Hurley et al. (2017)]{hurley17}Hurley, P.D., et al. 2017, MNRAS, 464, 885  

\bibitem[Jones et al. (2015)]{jones15}Jones, S.F., et al. 2015, MNRAS, 448, 3325 

\bibitem[Johnson et al. (2013)]{johnson13}Johnson, S.P., et al.  2013, MNRAS, 436, 2535 

\bibitem[Kelly et al. (2010)]{kelly10}Kelly, B.C., et al. 2010, ApJ, 719, 1315 

\bibitem[Laigle  et al. (2016)]{laigle16}Laigle, C.,  et al. 2016, ApJS, 224, 24 

\bibitem[Lonsdale et al.(2015)]{lonsdale15} 
Lonsdale, C.~J., Lacy, M., Kimball, A.~E., et al.\ 2015, \apj, 813, 45 

\bibitem[Ma\l{}ek et al. (2018)]{malek18}Ma\l{}ek, K., et al. 2018, AA, 620, 50

\bibitem[Marchesi et al. (2016)]{marchesi16}Marchesi, S., et al. 2016, ApJ, 817, 34

\bibitem[Mattila et al.(2012)]{mattila12} 
Mattila, S., Dahlen, T., Efstathiou, A., et al.\ 2012, \apj, 756, 111 

\bibitem[Mattila et al.(2018)]{mattila18} 
Mattila, S., P{\'e}rez-Torres, M., Efstathiou, A., et al.\ 2018, Science, 361, 482 

\bibitem[Mortlock et al. (2011)]{mortlock11}Mortlock, D.J., et al. 2011, Nature, 474, 616  

\bibitem[Noll et al.(2009)]{noll09} 
Noll, S., Burgarella, D., Giovannoli, E., et al.\ 2009, \aap, 507, 1793 

\bibitem[Pitchford et al.(2019)]{pitchford19} 
Pitchford, L.~K., Farrah, D., Alatalo, K., et al.\ 2019, \mnras, 487, 3130

\bibitem[Pilbratt et al.(2010)]{pilbratt10} 
Pilbratt, G.~L., Riedinger, J.~R., Passvogel, T., et al.\ 2010, \aap, 518, L1

\bibitem[Rafelski et al. (2012)]{rafelski12}Rafelski, M., et al. 2012, ApJ, 755, 89 

\bibitem[Riechers et al. (2014)]{riechers14}Riechers, D.A.,  et al. 2014, ApJ, 786, 31 

\bibitem[Ruiz et al.(2001)]{ruiz01} 
Ruiz, M., Efstathiou, A., Alexander, D.~M., et al.\ 2001, \mnras, 325, 995

\bibitem[Scoville et al. (2007)]{scoville07}Scoville, N., et al. 2007, ApJS, 172, 1 

\bibitem[Shirley et al.(2019)]{shirley19} 
Shirley, R., Roehlly, Y., Hurley, P.~D., et al.\ 2019, \mnras, 490, 634

\bibitem[Smolcic et al. (2017)]{smolcic17}Smolcic, V, et al. 2017, AA, 602, 1 

\bibitem[Spinoglio et al. (2017)]{spinoglio17}Spinoglio, L., et al. 2017, PASA, 34, 57 

\bibitem[Tsai  et al. (2015)]{tsai15}Tsai, C-W., et al. 2015, ApJ, 805, 90 

\bibitem[Tsai et al. (2018)]{tsai18}Tsai, C-W.,  et al. 2018, ApJ, 868, 15 

\bibitem[Verma et al.(2002)]{verma02} 
Verma, A., Rowan-Robinson, M., McMahon, R., et al.\ 2002, \mnras, 335, 574 

\bibitem[Vito  et al. (2018)]{vito18}Vito, F.,  et al. 2018, MNRAS, 473, 2378 

\bibitem[Wang (2018)]{wang18}Wang, F., 2018, ApJ, 869, L9 

\bibitem[Wu et al. (2015)]{wu15}Wu, X-B, et al. 2015, Nature, 518, 512 

\bibitem[Yang et al. (2020)]{yang20}Yang, J., et al. 2020, ApJL, 897, L14

\end{thebibliography}





\clearpage

 \begin{figure}
	\includegraphics[width=0.8\textwidth, angle=180]{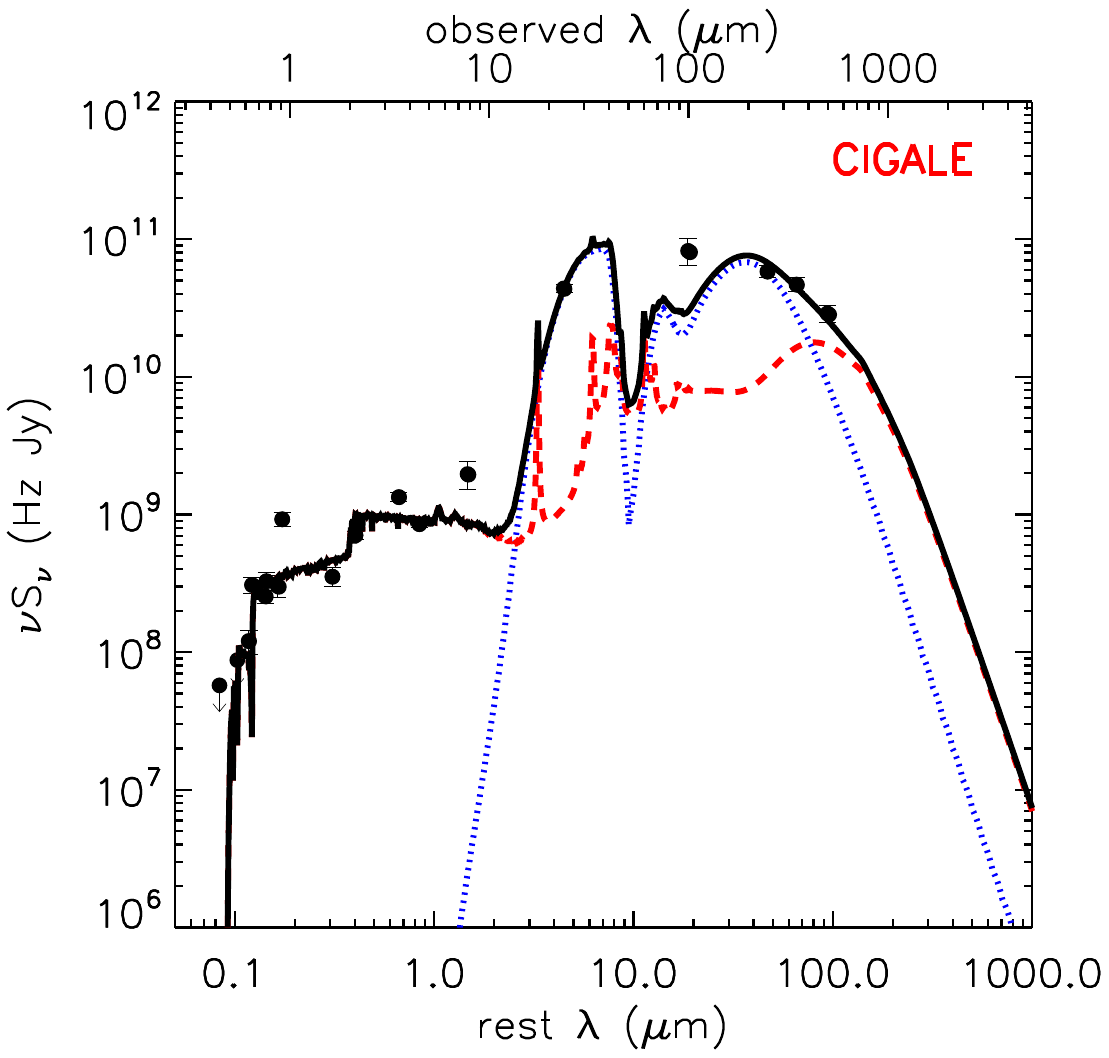}
	\vspace{-20pt}\par
    {\bf Figure S1} Best fit to the SED of HELP\_J100156.75+022344.7 with CIGALE.  The emission from the host galaxy is indicated by the red dashed line, the AGN torus emission by the blue dotted line and the total emission by the black solid line.
\end{figure}

\vfill\eject

 \begin{figure*}
	\centering
	\includegraphics[width=1.4\textwidth,angle=180]{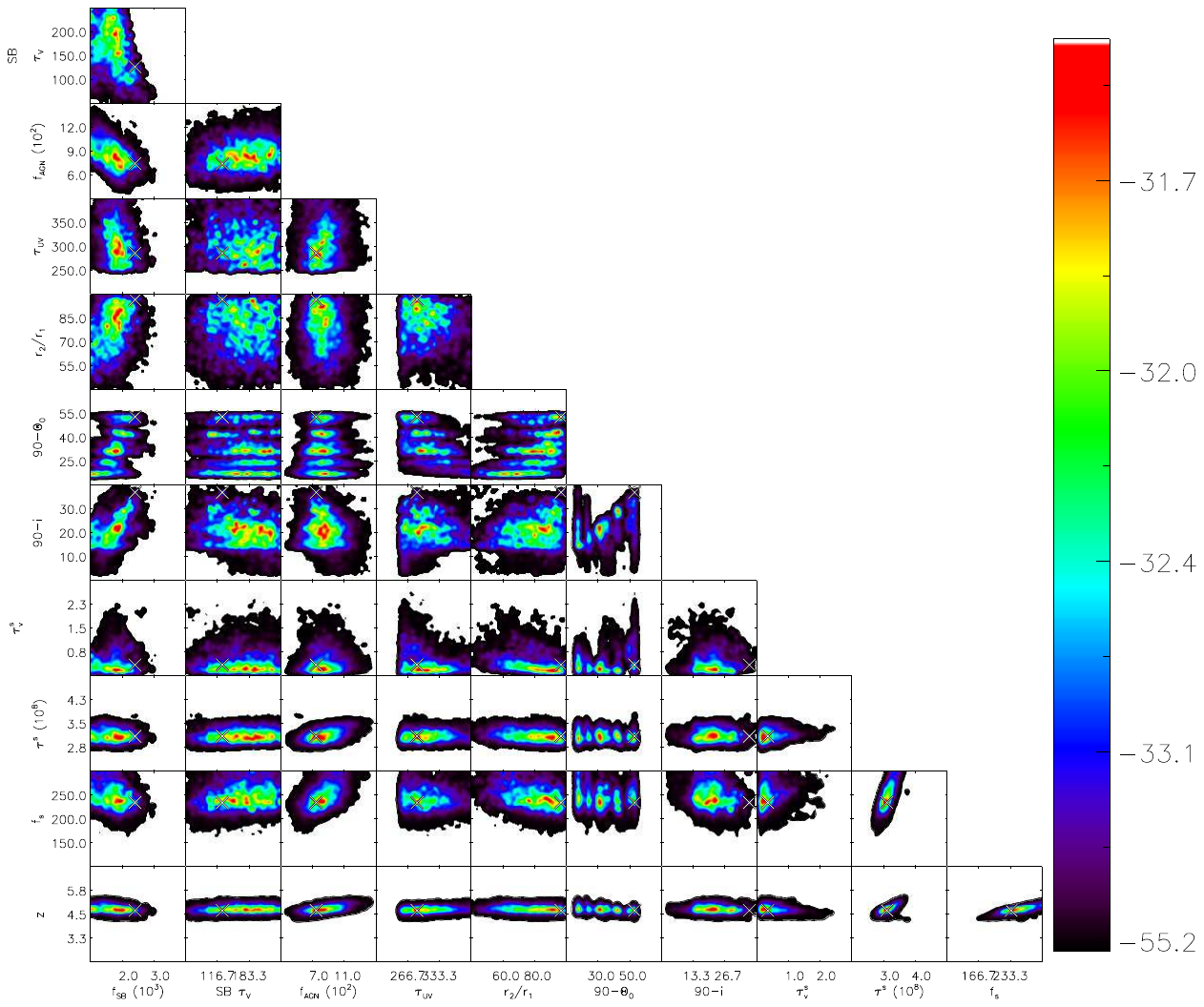}\par
    {\bf Figure S2} Plot showing the covariances between the parameters of the SED fit of HELP\_J100156.75+022344.7 with the CYGNUS models and using the photo-z option of SATMC. The X symbols mark the best fit. Also shown in color is the variation of the {log-likelihood} from its maximum of  -32 according to the color scheme on the right.
\end{figure*}

 \begin{figure*}
	\centering
	\includegraphics[width=1.4\textwidth,angle=180]{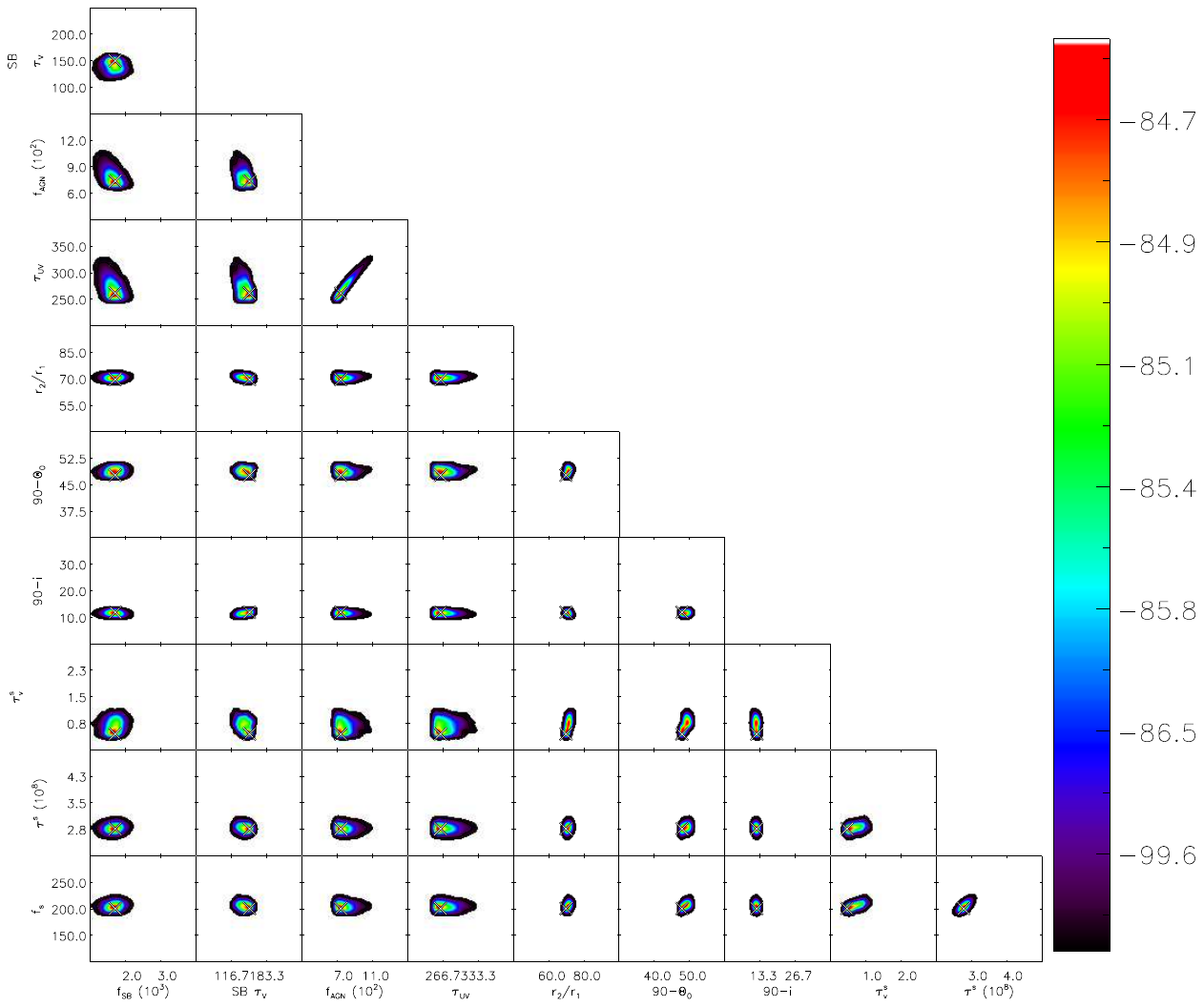}\par
    {\bf Figure S3} Same plot as Figure S2 showing the covariances between the parameters of the SED fit of HELP\_J100156.75+022344.7 with the CYGNUS models and $z$ fixed at 4.33.
\end{figure*}

 \begin{figure*}
	\centering
	\includegraphics[width=1.1\textwidth,angle=0]{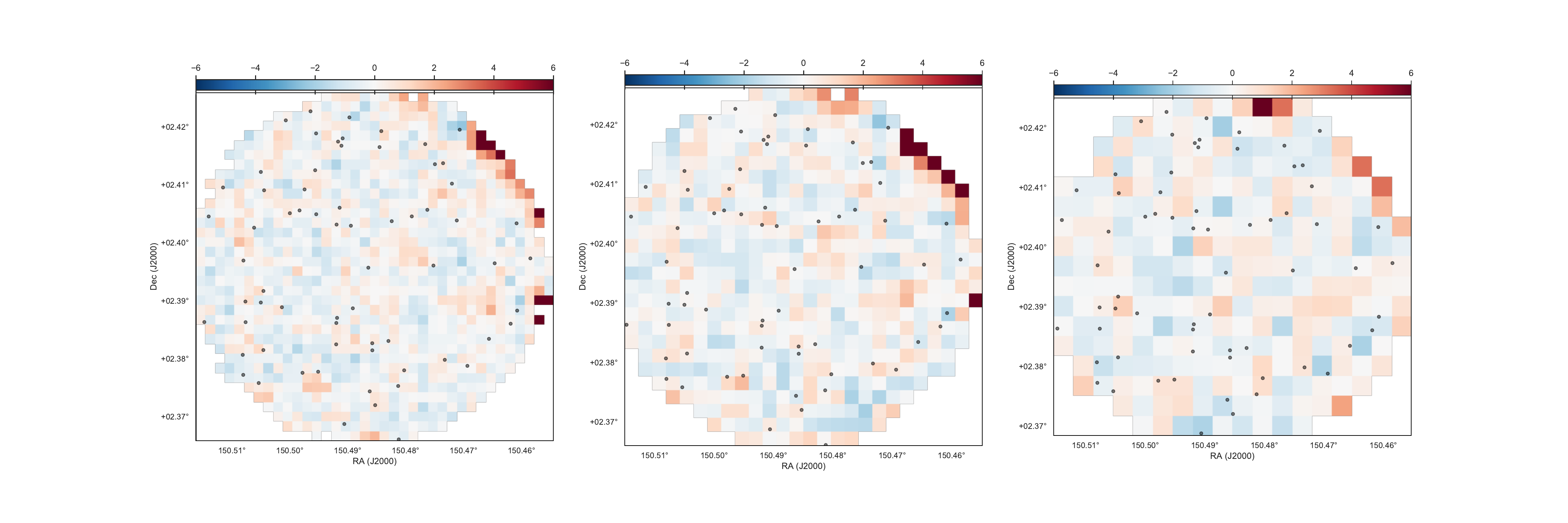}\par
    {\bf Figure S4} Plot showing the p-value maps at 250$\mu m$, 350$\mu m$ and 500$\mu m$ (from left to right). 
    The p-value maps are a form of Bayesian residual map. Each pixel in the p-value map is the effective sigma value of the true pixel flux in the distribution of model pixel fluxes from the posterior. Each pixel p-value is therefore a measure of how well the model accounts for the true pixel flux.
    The fact that all pixels have a p-value between -1 and 1$\sigma$ gives us confidence that the prior list used by XID+ \citep{hurley17} is accounting sufficiently for the map and important priors are not missed.
    The dots denote all the priors used by XID+.
\end{figure*}

\bsp	
\label{lastpage}
\end{document}